\documentclass[preprint, 3p, number, sort&compress]{elsarticle}
\usepackage{amsmath}
\usepackage{amsfonts}
\usepackage{latexsym}
\usepackage[pdftex]{color}
\usepackage{amsthm}
\usepackage{graphicx}
\usepackage{amssymb}
\usepackage{subfig}
\usepackage{float}
\usepackage{parskip}
\usepackage{textcomp}
\usepackage{array}
\usepackage{longtable}
\usepackage[colorlinks=true,citecolor=blue, linkcolor=blue,urlcolor=blue]{hyperref}
\usepackage{lipsum}

\newproof{pf}{\bf{Proof}}
\numberwithin{equation}{section}
\numberwithin{figure}{section}

\begin{document}

\begin{frontmatter}
\title{A difference-equation formalism for the nodal domains of separable billiards}

\author[rvt]{Naren Manjunath}
\author[rvt]{Rhine Samajdar}

\author[focal]{Sudhir R. Jain\corref{cor1}}
\ead{srjain@barc.gov.in}

\cortext[cor1]{Corresponding author, Phone: +91 222 559 3589}
\address[rvt]{\textsl{Indian Institute of Science, Bangalore 560012, India.}}
\address[focal]{\textsl{Nuclear Physics Division, Bhabha Atomic Research Centre, Mumbai 400085, India.}}

\begin{abstract}
Recently, the nodal domain counts of planar, integrable billiards with Dirichlet boundary conditions were shown to satisfy certain difference equations in [Ann. Phys. {\textbf{351}}, 1 (2014)]. The exact solutions of these equations give the number of domains explicitly. For complete generality, we demonstrate this novel formulation for three additional separable systems and thus extend the statement to all integrable billiards. 
\end{abstract}

\begin{keyword}
Integrable billiards \sep Nodal domains \sep Quantum chaos
\end{keyword}

\end{frontmatter}

\section{Introduction}
The classical billiard is a dynamical system consisting of a point particle moving freely in an enclosure, alternating between motion along a straight line and elastic reflections off the boundary, dictated by Snell's law \cite{Berry1981a, Eckhardt1984, Gutkin1986, Jain1992}. This sequence of specular reflections is captured by the billiard map, which completely describes the motion of the particle. These simple systems exhibit a wide range of dynamical behavior from order to chaos \cite{Berry1981b, Rabouw1981, Jain1995}, depending on the shape of their boundary. Classically speaking, an integrable billiard is defined to be one in which the number of constants of motion equals the number of degrees of freedom. A long-standing conjecture by Birkhoff states that among all billiards inside smooth convex curves, ellipses are characterized by integrability of the billiard map \cite{Birkhoff1927}. On the other hand, examples of their ergodic counterparts (for example, Bunimovich stadium \cite{Bunimovich1979}, dispersive Sinai billards \cite{Sinai1970}) are equally well-known \cite{Kozlov1991}.
 
Over the last two decades, the quantum analogues of these systems, quantum billiards, have been experimentally realized in gated, mesoscopic GaAs tables \cite{Berry1994}, microwave cavities \cite{Richter1999} and ultracold atom traps \cite{Milner2001, Friedman2001}. The eigenfunctions of these planar billiards organize themselves into regions, or domains, with positive and negative signs, often in remarkably complicated geometric shapes. Formally, such nodal domains may be defined as the maximally connected regions wherein the wavefunction does not change sign. Academic interest in various statistical measures pertaining to the number of nodal domains, $\nu$, was piqued \cite{Smilansky2007} in light of discoveries such as a new criterion for chaos in quantum mechanics \cite{Blum2002} and the presence of geometric information about the system in its nodal count sequences \cite{Jakobson2001}. Experimentally, nodal domains have also been the focus of much attention as documented in Refs.~\cite{Savytskyy2004, Hul2005, Kuhl2007}. Unfortunately, quantifying the nodal patterns is a major challenge since it is extremely hard to discern any order when $\nu$ is arranged in an ascending order of energy.

In principle, the problem seems (deceptively) straightforward---for each billiard of interest, we need only solve the Schr\"{o}dinger equation in appropriate coordinates, and count the domains as a function of the two quantum numbers $m$ and $n$ of the system. However, in order to arrive at a functional form for $\nu$, it was recently discovered \cite{Samajdar2014b} that it is more fruitful to analyze the differences $\Delta_{kn}\, \nu(m,n) =  \nu_{m+kn,n} - \nu_{m,n}, \,k \in \mathbb{N}$, instead. The proposition put forth therein was that for any integrable billiard in two dimensions, ``one of $\Delta_{kn} \nu\,(m,n) = \Phi(n) $ and $\Delta^2_{kn} \nu\,(m,n) = \Phi(n)\; \forall \, m, n$ holds for some $ \Phi: \mathbb{R} \rightarrow \mathbb{R}$, which is determined only by the geometry of the billiard.'' This difference-equation formulation, which proves to be of great practical utility in determining $\nu$ analytically (especially in the study of non-separable polygons, cf. \cite{Samajdar2014a}), was illustrated in \cite{Samajdar2014b} for a few separable and all non-separable, integrable polygonal shapes. The natural question that one then asks is: can \textsl{all} integrable, planar billiards be similarly characterized by  a difference equation in their nodal domain counts? The answer, as we show, is in the affirmative. 

For definiteness, we demonstrate that the statement can be generalized to \textsl{all} planar, integrable billiards, which includes both convex billiards, with smooth boundaries, and billiards in polygons. Since the two-dimensional Helmholtz equation is separable in only four coordinate systems---the Cartesian, polar, elliptic and parabolic coordinates \cite{Eisenhart1934, Eisenhart1948}---the additional geometries which must be considered are the elliptical billiard, the system of two confocal parabolas and various annular regions of the above. For all such separable systems, $k = 1$ and this may be regarded as a fingerprint of separability in the difference equations themselves.

\section{Circular annuli and sectors}
\begin{figure}[htb]
\begin{center}
\subfloat[]{\label{fig:1a}\scalebox{0.29}{\includegraphics{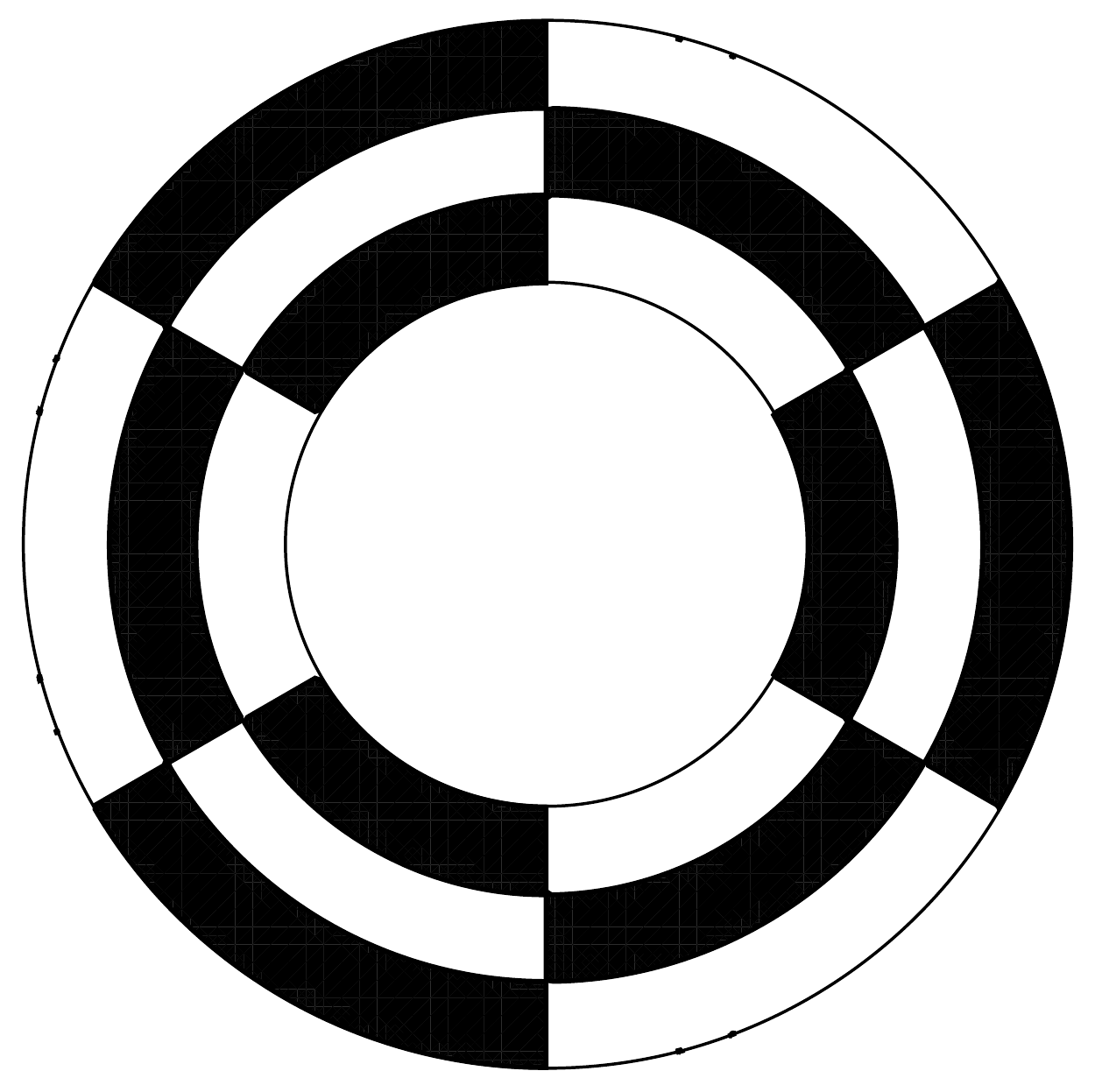}}}\quad
\subfloat[]{\label{fig:1b}\scalebox{0.29}{\includegraphics{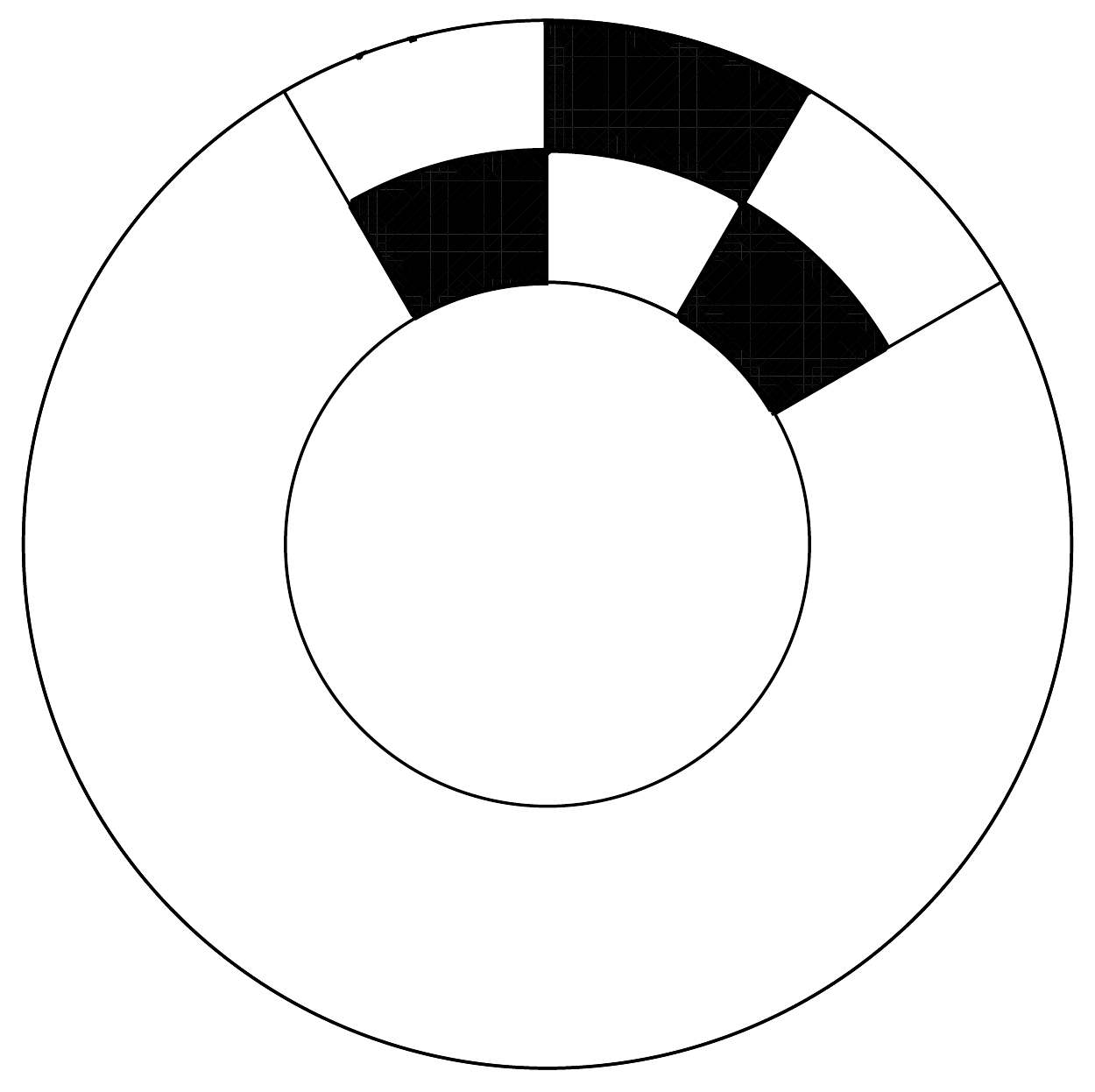}}}\quad
\subfloat[]{\label{fig:1c}\scalebox{0.29}{\includegraphics{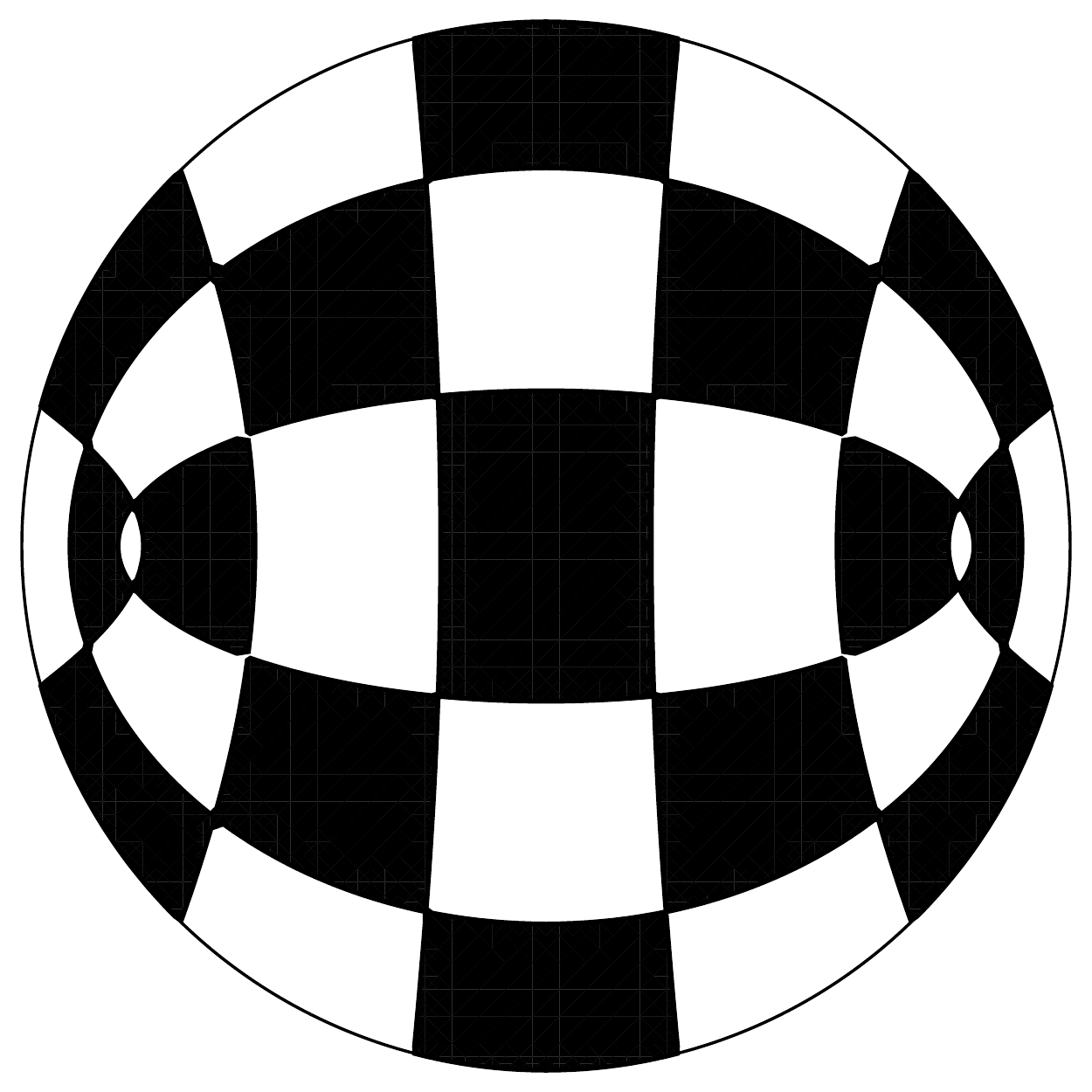}}}\quad 
\subfloat[]{\label{fig:1d}\scalebox{0.30}{\includegraphics[ trim={0cm 0.5cm 0.2cm 0cm}, clip]{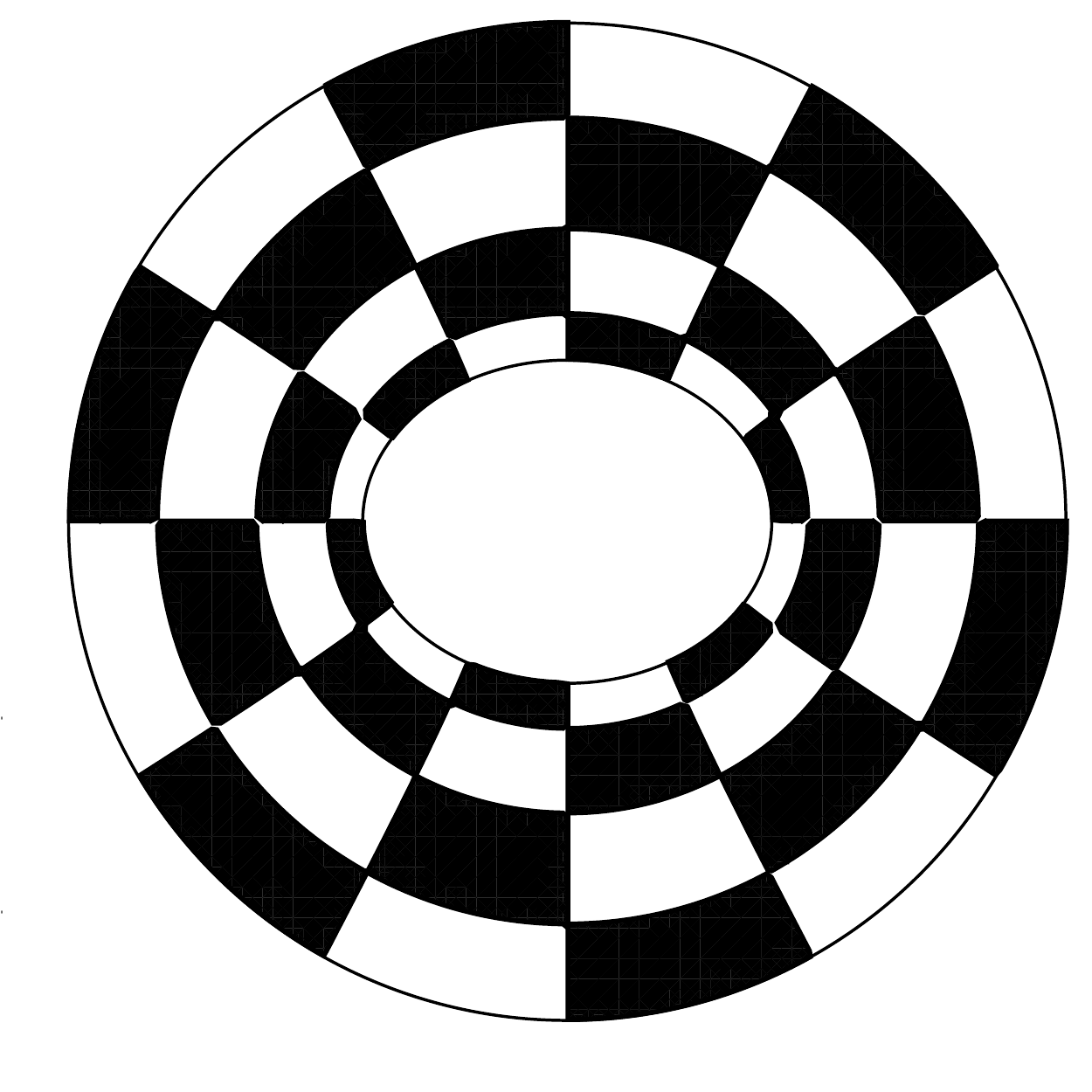}}}
\end{center}
\caption{\label{fig:AnnulusCircle} Nodal domains of $\psi_{m,n}$ on a) the circular annulus for $m=3$, $n=3$, where $m$ and $n$ are the quantum numbers for the angular and the radial variables, respectively, and on b) an annular sector of the circle, restricted in both the radial and angular coordinates ($r_2 = 2\, r_1$, $30^{\circ} \le \theta \le 120^{\circ}$) with $m=3$, $n=2$. The ratio of the radii $r_1$ and $r_2$ affects the relative areas of the nodal domains but not the nodal count. c) Nodal domains of the elliptic billiard in the (++) symmetry mode and d) for a confocal annular elliptic lake in the (- -) mode, corresponding to the state $r = 2$, $l=3$.}
\end{figure}

The circle is but a special case within the class of elliptical billiards and is described in two-dimensional polar coordinates by $\mathcal{D} = \{(r,\theta):0\le r \le R, 0\le \theta \le 2\pi \}$. The Helmholtz equation for this system, 
$r\, \partial_r (r\, \partial_r \psi) + \partial_{\theta}^2\, \psi + k^2 r^2 \psi = 0$, 
can be separated into radial and angular components, where the radial solution is a cylindrical Bessel function of the first kind, denoted by $J_m(kr)$, $m$ being the angular quantum number, and the angular function is simply $\exp (\mathrm{i} \,m \,\theta)$ \cite{Robinett1996, Robinett2003}. In Ref.~\cite{Samajdar2014b}, it was shown that for the circular billiard, $\Delta_{n} \nu(m,n) = \nu_{m+n,n} - \nu_{m,n} = 2 n^2$, which gives $\nu_{m,n} = 2\, m n$. Here, we study the annular regions of this billiard where the domain is restricted, first in the radial variable and then, in both the radial and angular variables. For the annulus where $\theta \in [0,2\pi]$, the quantum number $m$ so obtained is integral and thus the general solution to the radial Helmholtz equation is a linear combination of the $m^{\mbox{\scriptsize{th}}}$-order Bessel functions of the first and second kind, $J_m(kr)$ and $Y_m(kr)$. Assuming that the annular region is enclosed within the radii $r_1$ and $r_2$, the boundary conditions $\psi = 0$ for $r = r_1, r_2$ suggest the form
\begin{equation}
\psi_{\,m,n}(r,\theta) = A\,  \big[ J_m(k_n r)\, Y_m(k_n r_1) - J_m(k_n r_1)\, Y_m(k_n r) \big] \cos(m \pi \theta ),
\end{equation}
where $A$ is a normalization constant and $k_n$ is the value of $k$ such that $r_2$ is the $n^{\mbox{\scriptsize{th}}}$ zero of the radial solution in the domain of its definition (excluding the zero at $r_1$). Adopting these conventions for $m$ and $n$, the recurrence relation satisfied by the nodal domains is $\nu_{m+n,n} - \nu_{m,n} = 2 n^2$, and on scrutinizing individual cases, the formula $\nu_{m,n} = 2\, m n$ is obtained. These results are identical to those obtained for the full circle and the corresponding examples have been illustrated in Fig.~\ref{fig:1a}. On the other hand, for regions where the domain is restricted in the radial as well as the angular variable, the Dirichlet boundary conditions become analogous to those imposed on a rectangular billiard, which can be argued as follows. In the cases where $\theta$ runs from $0$ to $2 \pi$, the requirement of periodicity in $\theta$ forces the solution to the angular equation to have $2m$ zeros in $[0,2 \pi]$. Contrarily, if $\theta$ is restricted, the need for periodicity no longer holds and the number of zeros of the angular equation will just be $m$. Hence, one expects $\nu_{m,n} = m n$ and $\nu_{m+n,n} - \nu_{m,n} = n^2$, which is suggestively reminiscent of the corresponding relations for the rectangular billiard. Upon examination, this is indeed found to be true, as has been shown in Fig.~\ref{fig:1b}, although we do not explicitly present the calculations for this case. Moreover, it will be seen that for all separable systems, when the boundary of the billiard is restricted in both coordinates such that the endpoints do not coincide (as in the case of the rectangle), $\nu_{m,n}$ is always $m n$. 

\section{Ellipses and elliptical annuli}
The elliptical billiard is simply the region $\mathcal{D} = [0,\rho_{max}] \times [0,2 \pi)$ in elliptic coordinates, $(\rho, \phi)$, where lines of constant $\rho$ are confocal ellipses while those of constant $\phi$ belong to a family of confocal hyperbolas. The Helmholtz equation again splits into radial and angular components, which satisfy the Mathieu equations: 
\begin{equation}
\label{Mathieu}
\partial_{\rho}^2 F - \big(\,\lambda - c^2 \cosh^2 \rho\,\big)\,F = 0, \quad
\partial_{\phi}^2 G + \big(\,\lambda - c^2 \cos^2\phi\,\big)\,G = 0,
\end{equation}
where $\psi(\rho,\phi) = F(\rho)\, G(\phi)$ and $\psi(\rho_{max}, \phi) \equiv 0$. The solutions to these equations are the radial and angular Mathieu functions \cite{Vega2003} and are constructed so as to be symmetric (+) or antisymmetric (-) with respect to the $x$ and $y$ axes (refer Fig.~\ref{fig:1c}). The quantum numbers $r$ and $l$ are defined so as to be consistent with the Einstein-Brillouin-Keller (EBK) quantization for the symmetry reduced system \cite{Waalkens1997}. For the elliptical annulus, defined as the regions $\mathcal{D} = \{(\rho,\phi): \rho_{min} \le \rho \le \rho_{max}\}$, we follow a procedure analogous to the case of the circular annuli. The wavefunction we seek is a linear combination of the even and odd Mathieu functions that are independent solutions to the Mathieu equations given above. By convention, the radial quantum number $r$ is defined to be the number of zeros of the radial equation within the open interval $(\rho_{min},\rho_{max})$. The angular quantum number $l$ is defined as the number of zeros of the angular Mathieu equation in $(0,\pi/2)$. Careful inspection shows that each mode possesses its individual difference equation and $\Delta_{l}\, \nu_{r,l} $ is obtained and summarized in Table~\ref{Table:Ellipse} for the four symmetry classes, along with the corresponding expressions for $\nu_{r,l}$ derived therefrom. As opposed to the circular well, the exact expressions for $\nu_{r, l}$ differ between the elliptic billiard and its annular slices.
\bgroup
\def\arraystretch{1.5}
\begin{table}[htb]
\centering
{
\bgroup
\small
\setlength{\tabcolsep}{19.65pt}
\begin{tabular}{c r r| r r} \hline
\multicolumn{1}{c}{Symmetry Mode} &\multicolumn{1}{c}{$\Delta_{l}\, \nu_{r,l} $} &\multicolumn{1}{c|}{$\nu_{r,l}$} &\multicolumn{1}{c}{$\Delta_{l}\, \nu_{r,l} $} &\multicolumn{1}{c}{$\nu_{r,l}$}  \\ \hline
$++$ &$4l^2$&$2 l(2r+1) + 1$&$4l^2$&$4l(r+1)$ \\
$+-$ &$4l^2 + 2l$&$2(2l+1)(r+1)$&$4l^2 + 8l$&$2(2l+1)(r+1)$ \\
$-+$ &$4l^2 + 2l$&$(2l+1)(2r + 1) + 1$&$4l^2 + 8l$&$2(2l+1)(r+1)$ \\
$--$ &$4l^2 + 4l$&$4(l+1)(r+1)$&$4l^2 + 4l$&$4(l+1)(r+1)$ \\ \hline
\end{tabular}
}
\egroup
\caption{\label{Table:Ellipse}A summary of the difference equations and the formulae for nodal domain numbers of the elliptic billiard (left panel) and the elliptic annulus with $\rho_{min} = 1$ and $\rho_{max} = 2$ (right panel). When $l=0$, the difference equation still holds whereby $\Delta_{l}\, \nu_{r,0} = 0$, although the particular solution has to be now modified. The expressions for the first difference and $\nu_{r,l}$ in the annulus are, of course, completely general and independent of the (arbitrary) limits of $\rho$. The theoretical predictions tabulated here are verified by Figs.~\ref{fig:1c} and \ref{fig:1d}. }
\end{table}
\egroup

\section{The confocal parabolic billiard}

The system of two confocal parabolae had never even been studied in the context of billiard motions until very recently \cite{Fokicheva2014}, and is undocumented from a quantum mechanical perspective---we therefore present our analysis in some detail. The parabolic coordinates $(\tau,\,\sigma)$ are defined by $x = \tau \sigma $ and $y = (\tau^2 - \sigma^2)/2$ with $\sigma \ge 0$. In this system, lines of constant $\sigma$ and $\tau$ form upward- and downward-facing families of parabolae, respectively, with their common focus at the origin. The lines $\sigma = 0$ and $\tau = 0$ correspond to the positive and negative branches of the $y$-axis, respectively. The billiard under consideration is the region 
$\mathcal D = \{(\tau,\sigma) : \lvert \tau \rvert\le 1 \,$ and $\, 0 \le \sigma \le 1\}$. The Helmholtz equation in parabolic coordinates is \cite{Morse1953, Moon1988}
\begin{equation}
\label{Parabola}
\big(\partial_{\sigma}^2 + \partial_{\tau}^2\big)\, \psi + k^2\, (\sigma^2 + \tau^2)\, \psi = 0, 
\end{equation}
with $\psi = 0 \,$ on $\, \partial \mathcal D$. Using the ansatz $\psi(\tau,\sigma) = T(\tau) S(\sigma)$, Eq.~\eqref{Parabola} splits into
\begin{equation}
T^{\prime\prime}(\tau) + (k^2 \tau^2 + m)\, T = 0, \quad
S^{\prime\prime} (\sigma) + (k^2 \sigma^2 - m)\, S = 0,
\end{equation}
where $m$ is a separation constant. Employing a change of variables to $u = \sqrt{2k}\sigma$, $v = \sqrt{2k}\tau$ and defining functions $H$, $G$ such that
$ G(v) = T(\tau)$, $H(u) = S(\sigma) $ leads to the standard form 
\begin{equation}
G^{\prime\prime}(v) + \bigg(\frac{v^2}{4} + a\bigg)\, G = 0, \quad  
H^{\prime\prime} (u) + \bigg(\frac{u^2}{4} - a\bigg)\, H = 0,
\end{equation}
where $a$ is to be determined. Odd and even solutions to the equation for $G$ (and hence for $H$) can be expressed in terms of the Kummer hypergeometric functions $M(a,b,z)$. Specifically, two real and independent normalized solutions of $G$ are \cite{Abramowitz1970}
\begin{equation}
V_a(v) = e^{-\mathrm{i}v^2/4}\, M\bigg(-\frac{a}{2} + \frac{1}{4},\frac{1}{2},\frac{\mathrm{i}v^2}{2}\bigg), \quad
U_a(v) = \frac{1+i}{\sqrt{2}} v\, e^{-\mathrm{i}v^2/4}\, M\bigg(-\frac{a}{2} + \frac{3}{4},\frac{3}{2},\frac{\mathrm{i}v^2}{2}\bigg),
\end{equation}
whence the general solution of $T$ is a linear combination of $U_a$ and $V_a$. Similar considerations ($a$ being replaced by $-a$) hold for $S$ as well. In addition to the boundary conditions $T(\pm1) = 0$, $S(1) = 0$, $\psi$ must be continuous and differentiable across $\sigma = 0$ (the + $y$-axis) where $\tau$ changes discontinuously. This is possible in two ways: either $T$ can be even or $S$ can vanish on the $\sigma = 0$ line; in both cases, $S$ is either odd or even. If $\psi$ is zero on the positive $y$-axis, it must also vanish on the negative $y$-axis in order for it to be physically meaningful. This condition forces that $T$ and $S$ are either both even or both odd. We look at the two cases individually below.

\subsection{Case 1: $S$ and $T$ even}
Under these circumstances, $T = V_a (\sqrt{2k}\,\tau)$ and $S = V_{-a} (\sqrt{2k}\,\sigma)$. We require $T(1) = S(1) = 0$ and this is possible only for certain discrete values $a_{m,n}$ of $a$, where $m$ and $n$ are natural numbers such that the $m^{\mbox{\scriptsize{th}}}$ root of $T$ and the $n^{\mbox{\scriptsize{th}}}$ root of $S$ are coincident and equal some real number $w_{m,n}$. The wavenumber $k$, and consequently, the energy eigenvalues, are obtained by setting $2k = w_{m,n}^2$. Upon plotting the nodal domains of this function, depicted in Fig.~\ref{fig:3a}, we discover 
\begin{equation}
\Delta_n \nu (m, n) = \nu_{m+n,n} - \nu_{m,n} = 2n^2 - n \qquad
\mbox{and}\qquad  \nu_{m,n} = 2\,m n - m - n +1.
\end{equation}

\subsection{Case 2: $S$ and $T$ odd}
A similar analysis for this situation yields for the first difference $\nu_{m+n,n} - \nu_{m,n} = 2n^2$ whereupon, by vetting particular values of $\nu$, $\nu_{m,n} = 2\, m n$. 

\begin{figure}[htb]
\begin{center}
\subfloat[]{\label{fig:3a}\scalebox{0.29}{\includegraphics{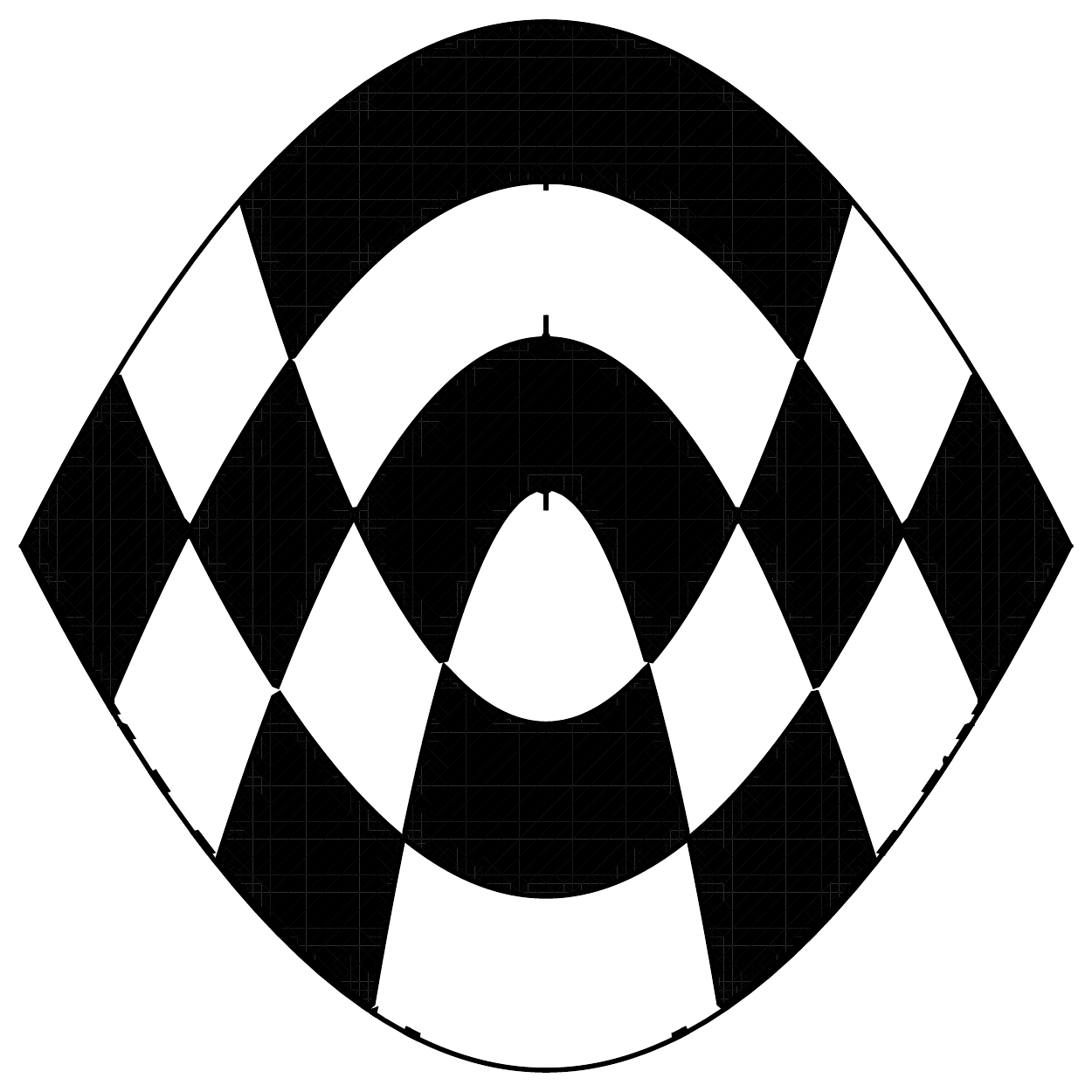}}}\quad
\subfloat[]{\label{fig:3b}\scalebox{0.29}{\includegraphics{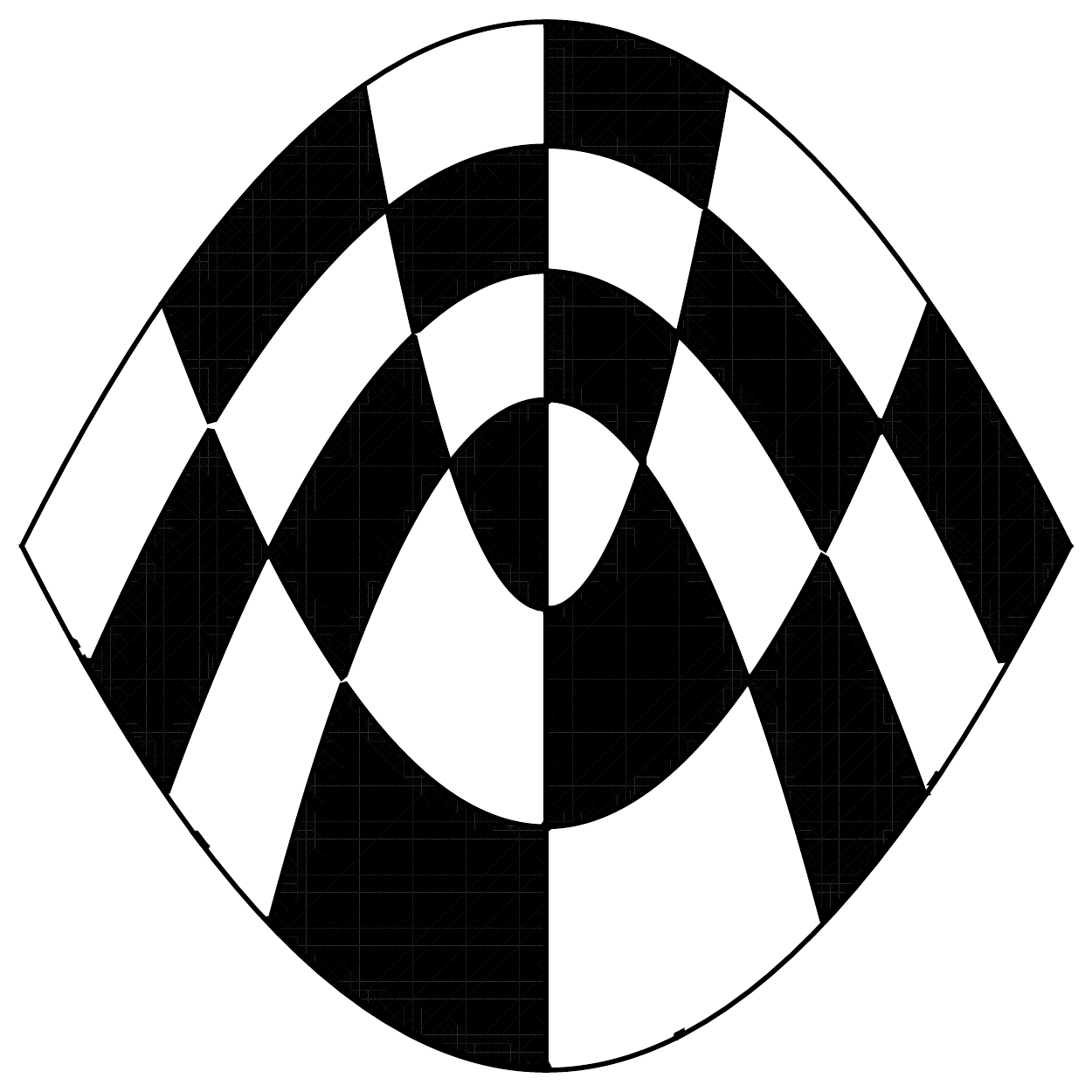}}}\quad
\subfloat[]{\label{fig:3c}\scalebox{0.29}{\includegraphics[width=0.80\textwidth,height=5.025in]{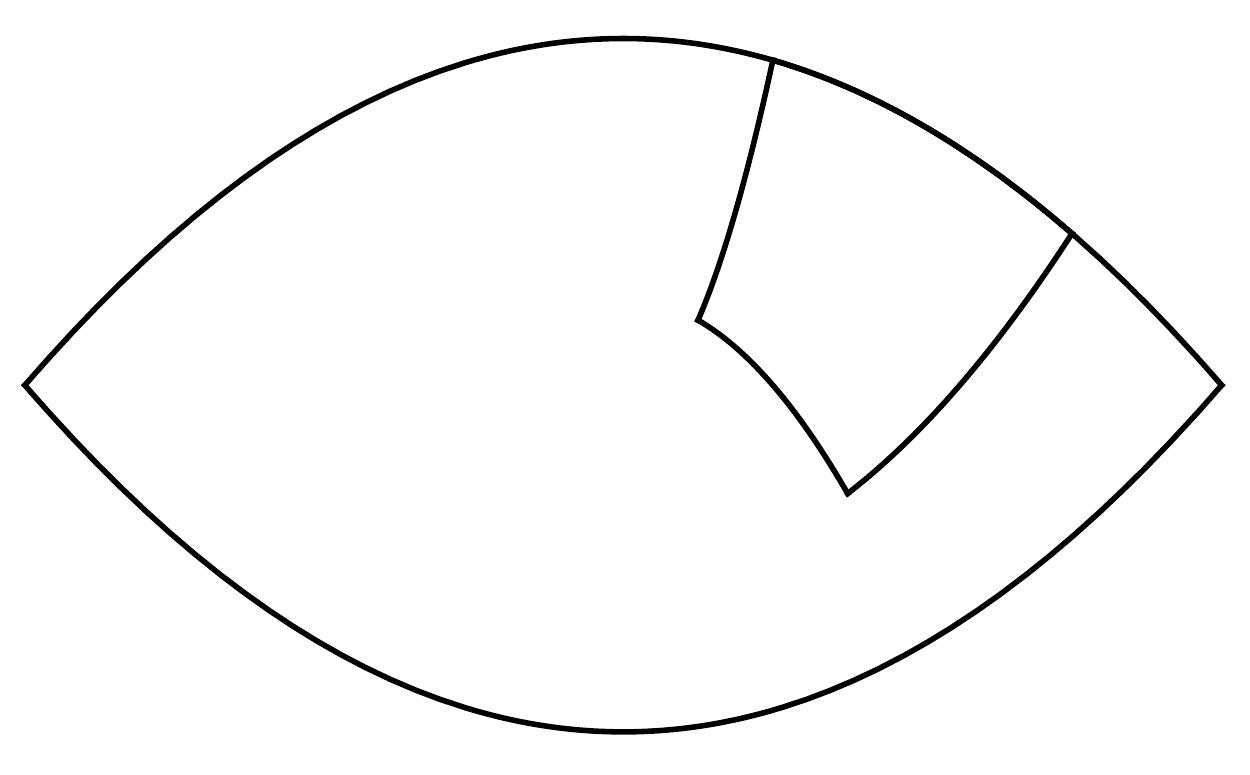}}}\quad
\subfloat[]{\label{fig:3d}\scalebox{0.29}{\includegraphics[width=0.75\textwidth,height=5.025in]{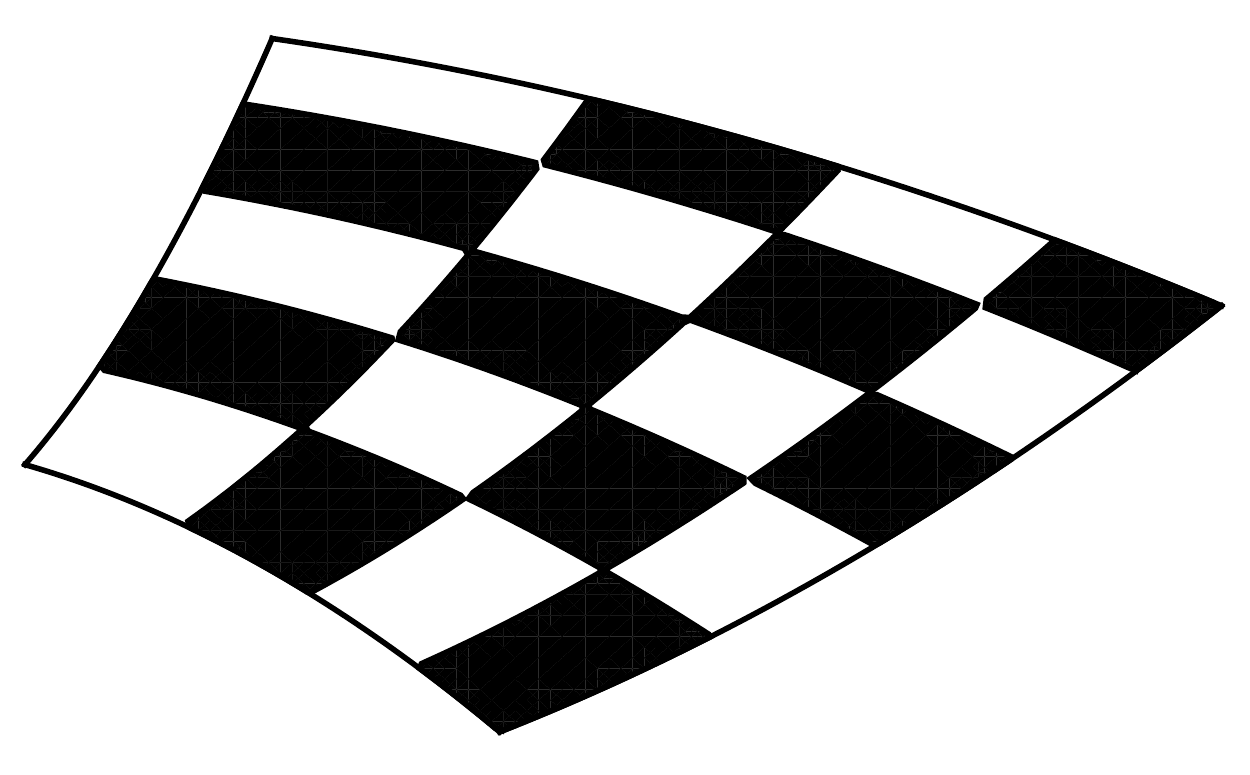}}}
\end{center}
\caption{\label{fig:Parabola}Nodal domains of the confocal parabolas showing $\psi_{4,3}$ in two cases, where the functions $S$ and $T$, whose product equals $\psi$, are a) both even and b) both odd. The nodal lines are themselves parabolas whose common focus is at the origin. For an annular segment of this system, the region of interest is demarcated in the left panel (c) with the original parabolas for comparison, while the figure on the right (d) displays $\psi_{\,5,4}$ on this tract.}
\end{figure}

The analysis of the parabolic annuli is fundamentally the same as for the other annular cases discussed earlier but the situation is computationally more involved, because $a$ and $k$ have to be determined simultaneously. We find that, as before, $\Delta_{n} \nu_{m,n} = n^2$ and that $\nu_{m,n} = mn$.  An example of such an annular domain is displayed in Fig.~\ref{fig:3c} for convenience of visualization. It is trivial to notice that when either of $m$ or $n$ is zero, there are no intersections of nodal lines and thus, $\nu_{m,0} = \nu_{\,0,n} = 1$, which includes the case $\nu_{\,0,0}$.

To summarize, data of actual nodal counts illustrating our methodology in the case of three separable billiards has been presented in Table~\ref{SepTable}. The existence of such a recurrence relation among members of the same equivalence class, defined by $m$ (mod $n$), is in complete agreement with our initial proposition. 

\bgroup
\def\arraystretch{1.5}
\begin{table}[htb]
\centering
{
\bgroup
\small
\setlength{\tabcolsep}{5.5pt}
\begin{tabular}{cccrcrcrcrc} \hline
&&& \multicolumn{2}{c}{Circular Annulus} &\multicolumn{2}{c}{Ellipse (++)} &\multicolumn{2}{c}{Elliptical Annulus (++)} & \multicolumn{2}{c}{Confocal Parabolas} \\ 
$m \equiv l$&$n \equiv r$&$m$ mod $n$&$\nu_{m,n}$ & $\Delta_{n}\nu_{m,n} $ & $\nu_{m,n}$ & $\Delta_{n}\nu_{m,n} $ &$\nu_{m,n}$ & $\Delta_{n}\nu_{m,n} $ & $\nu_{m,n}$ & $\Delta_{n} \nu_{m,n}$ \\ \hline
5&4&1&40&--&91&--&100&--&32&--\\
9&4&1&72&32&163&72&180&80&60&28\\
13&4&1&104&32&235&72&260&80&88&28\\
17&4&1&136&32&307&72&340&80&116&28\\ 
21&4&1&168&32&379&72&420&80&144&28\\ \hline
\end{tabular}
}\\[\baselineskip]
\egroup
\caption{\label{SepTable}An example of the procedure for determining the difference equation in $\nu_{m,n}$ for the three separable billiards discussed above. A particular $n$ is fixed and $\nu_{m,n}$ is tabulated for a sequence of values of $m$ from the same equivalence class $m \mod n$. The constancy of the first difference of $\nu_{m,n}$ for the wavefunctions belonging to the same class is easily seen and implies that $\nu_{m,n}$ is linear in $m$. By plotting similar tables for a few low-lying values of $n$, $\Delta_{n} \nu_{m,n}$ is determined as a function of $n$. }
\end{table}
\egroup

\section{Conclusion}
Two-dimensional quantum billiards have been very important for studying the manifestations of chaos in quantum systems \cite{Reichl2004}. Historically, testing a quantum system for integrability has proved to be a much harder problem than its classical counterpart.  Efforts in this direction have ranged across systematically searching for quantum integrals of motion using Moyal brackets \cite{Hietarinta1982, Hietarinta1984}, constructing hypothetical second invariants \cite{Peres1984}, finding suitable Lax pairs \cite{Ganoulis1987}, and perhaps most ubiquitously, examining the statistical properties of the quantised energy spectrum \cite{McDonald1979}. In particular, soon after Percival's conjecture \cite{Percival1973}, it was realized  that the level-spacing distributions of integrable systems, harmonic oscillators notwithstanding, exhibit Poisson-like behavior \cite{Berry1977}. However, this analysis had its own shortcomings; to wit, a histogram of nearest-neighbor spacings for the simplest integrable billiard, the rectangle, could not be fit to a Poisson curve with good $\chi^2$ confidence even with $10^5$ levels \cite{Casati1985}! Moreover, one may not always have enough level spacings to ensure the statistical significance of a fit of the Poisson or Wigner-Dyson distributions, as is indeed the case with the confocal parabolic billiard, for which the determination of $w_{m,n}$ is computationally quite laborious. 

In this article, we therefore offer an alternative and unequivocal test of integrability for a broad class of systems, the planar separable billiards, premised on the existence of difference equations in their nodal domain counts. Coupled with the fact that the only integrable but non-separable billiards in two dimensions---the right-angled isosceles, the equilateral and the $30^{\circ}-60^{\circ}-90^{\circ}$ hemiequilateral triangle \cite{Schachner1994, Kaufman1999}---are already known to exhibit such a property \cite{Samajdar2014b}, this completes a comprehensive description of the recurrence relations in \textsl{all} planar, integrable billiards and hints at a signature of integrability, hidden in the nodal domain counts alone. Furthermore, our framework also prescribes a convenient methodology to obtain analytical expressions for the total number of domains of all such billiards in two dimensions, thereby mathematically quantifying the geometric nodal patterns and establishing a much-needed foundation for further statistical investigations. The generalization to three dimensions is expected to be along the same lines.

% References

\bibliography{Jabref/References}

\end{document}